\documentclass[aps,showpacs,preprintnumbers,amsmath,amssymb,nofootinbibt,twocolumn]{revtex4}
\usepackage{graphicx}

\begin{document}


\title{Two-fluid dark matter models}
\author{Tiberiu Harko}
\email{harko@hkucc.hku.hk} \affiliation{Department of Physics and
Center for Theoretical and Computational Physics, The University
of Hong Kong, Pok Fu Lam Road, Hong Kong, P. R. China}

\author{Francisco S.~N.~Lobo}
\email{ flobo@cii.fc.ul.pt}
\affiliation{Centro de Astronomia e Astrof\'{\i}sica da
Universidade de Lisboa, Campo Grande, Ed. C8 1749-016 Lisboa,
Portugal}

\date{\today}

\begin{abstract}

We investigate the possibility that dark matter is a mixture of two non-interacting perfect fluids, with different four-velocities and thermodynamic parameters. The two-fluid model can be described as an effective single anisotropic fluid, with distinct radial and tangential pressures. The basic equations describing the equilibrium structure of the two-fluid dark matter model, and of the tangential velocity of test particles in stable circular orbits, are obtained for the case of a spherically symmetric static geometry. By assuming a non-relativistic kinetic model for the dark matter particles,  the density profile and the tangential velocity of the dark matter mixture are obtained by numerically integrating the gravitational field equations. The cosmological implications of the model are also briefly considered, and it is shown that the anisotropic two-fluid model isotropizes in the large time limit.

\end{abstract}

\maketitle

\section{Introduction}

Cosmological observations provide compelling evidence that about 95\% of the
content of the Universe resides in two unknown forms, denoted
dark matter and dark energy, respectively. The former resides in bound objects as
non-luminous matter, and the latter in the form of a zero-point energy that
pervades the whole Universe \cite{PeRa03}. Dark matter is thought to be
composed of cold neutral weakly interacting massive particles, beyond those
existing in the Standard Model of Particle Physics, and not yet detected in
accelerators or in direct and indirect searches. There are many
possible candidates for dark matter, the most popular ones being the axions
and the weakly interacting massive particles (WIMP) (for a review of the
particle physics aspects of dark matter see \cite{OvWe04}). Their
interaction cross section with normal baryonic matter, while extremely
small, are expected to be non-zero and we may expect to detect them
directly.  Scalar fields or other long range
coherent fields coupled to gravity have also been used intensively to model
galactic dark matter \cite{scal}. Recently, the possibility that the galactic dynamics of
massive test particles may be understood in the context of modified theories of gravity,
without the need for dark matter, was also explored~\cite{dmmodgrav}.

Despite these important achievements, at galactic scales $\sim 10$ kpc, the $
\Lambda $CDM model meets with severe difficulties in explaining the observed
distribution of the invisible matter around the luminous one. In order for galactic gas and
stars to attain a constant velocity in the outer regions, the density profile of the dark
matter must fall approximately with $r^{-2}$ in the respective regions. These profiles resemble the density structure of an isothermal and self-gravitating
system of particles, characterized by a constant velocity dispersion. In fact, $N$-body simulations, performed in this scenario, predict that bound halos
surrounding galaxies must have very characteristic density profiles that
feature a well pronounced central cusp \cite{Navarro:1994hi}. On the observational
side, high-resolution rotation curves show, instead, that the actual
distribution of dark matter is much shallower than the above, and it
presents a nearly constant density core \cite{Burkert:1995yz}.

The possibility that dark matter could be described by a fluid with non-zero effective pressure was also considered in the physical literature \cite{pres}. In particular, in \cite{Sax} it was assumed that the equation of state of the dark matter halos is polytropic. The fit with a polytropic dark halo improves the velocity dispersion profiles.

Due to their importance in astrophysics and cosmology, two-fluid models have been extensively investigated in the physical literature. In particular, dust shells of finite size, moving with different
four-velocities and forming a common region, where the two dust components exist simultaneously, were considered in \cite{Haag1} for the case of a Lemaitre-Tolman  geometry. To model voids a spherically symmetric
dust with different energy density values has been matched by an intermediate region. By introducing two dust components the boundaries no longer need to be comoving. The two-component dust model was generalized in \cite{Haag2} by introducing an additional homothetic vector.  The obtained metrics are regular, except for a big bang singularity and in the limit of large times they can be given by an asymptotic expansion of the metric coefficients. Using this family of solutions, voids whose edge is not comoving in comparison with its surroundings can be modeled in geometric terms.

In this work, we consider the possibility that dark matter may be modeled as a mixture of two non-interacting perfect fluids, with different four-velocities. This configuration is formally equivalent to a single anisotropic fluid \cite{Le,He}. We obtain the basic equations describing the structure of this system, namely, the mass continuity and the hydrostatic equilibrium equations, as well as the equation giving the tangential velocity of test particles in stable circular orbits. For the dark matter we adopt a non-relativistic kinetic description, with the fluid pressure proportional to the density of the fluid and to the velocity dispersion of the constituent particles. By assuming that the two fluids have the same density, the dark matter profile and the tangential velocity are obtained by numerically integrating the field equations. Some cosmological implications of the model  are also discussed, and it is shown that in the large time limit the initially anisotropic homogeneous two-fluid dark matter distribution ends in an isotropic phase.

\section{Dark matter as a mixture of two perfect fluids}

We start our study of dark matter by assuming that it consists of a mixture
of two perfect fluids, with energy densities and pressures $\rho _{1}$, $%
P_{1}$ and $\rho _{2}$, $P_{2}$, respectively, and with four velocities $%
U^{\mu }$ and $W^{\mu }$, respectively. The fluid is described by the total
energy-momentum tensor $T^{\mu \nu }$, given by
\begin{eqnarray}
T^{\mu \nu }&=&\left( \rho _{1}+P_{1}\right) U^{\mu }U^{\nu }-P_{1}g^{\mu \nu
}+\nonumber\\
&&\left( \rho _{2}+P_{2}\right) W^{\mu }W^{\nu }-P_{2}g^{\mu \nu }.
\label{emtensor}
\end{eqnarray}

The four-velocities are normalized according to $U^{\mu }U_{\mu }=1$ and $%
W^{\mu }W_{\mu }=1$, respectively. In the present paper we use the natural system of units with $c=G=1$. The study of the physical systems
described by an energy-momentum tensor having the form given by Eq.~(\ref
{emtensor}) can be significantly simplified if we cast it into the standard
form of perfect anisotropic fluids. This can be done by means of the
transformations \cite{Le,He}
\begin{eqnarray}
U^{\mu }&\rightarrow &U^{\ast \mu }=U^{\mu }\cos \alpha +\sqrt{\frac{\rho
_{2}+P_{2}}{\rho _{1}+P_{1}}}W^{\mu }\sin \alpha ,  \label{te1}\\
W^{\mu }&\rightarrow &W^{\ast \mu }=W^{\mu }\cos \alpha -\sqrt{\frac{\rho
_{1}+P_{1}}{\rho _{2}+P_{2}}}U^{\mu }\sin \alpha ,  \label{te2}
\end{eqnarray}
representing a ``rotation'' of the velocity four-vectors in the $\left(
U^{\mu },W^{\mu }\right) $ velocity space, which leave the quadratic form $\left(
\rho _{1}+P_{1}\right) U^{\mu }U^{\nu }+\left( \rho _{2}+P_{2}\right) W^{\mu
}W^{\nu }$ invariant. Thus, $T^{\mu \nu }\left( U,W\right) =T^{\mu \nu
}\left( U^{\ast },W^{\ast }\right) $. Next, we choose $U^{\ast \mu }$ and $W^{\ast \mu }$ such
that one becomes timelike, while the other is spacelike. Therefore the two vectors satisfy the condition
$U^{\ast \mu }W_{\mu }^{\ast }=0$.
With the use of Eqs.~(\ref{te1})-(\ref{te2}) and $U^{\ast \mu }W_{\mu }^{\ast }=0$ we obtain the rotation angle as
\begin{equation}\label{alpha}
\tan 2\alpha =2\frac{\sqrt{\left( \rho _{1}+p_{1}\right) \left( \rho
_{2}+p_{2}\right) }}{\rho _{1}+p_{1}-\left( \rho _{2}+p_{2}\right) }U^{\mu
}W_{\mu }.
\end{equation}
By defining the quantities
$V^{\mu }=U^{\ast \mu }/\sqrt{U^{\ast \alpha }U_{\alpha }^{\ast }}$,
$\chi ^{\mu }=W^{\ast \mu }/\sqrt{-W^{\ast \alpha }W_{\alpha }^{\ast }}$,
$\varepsilon =T^{\mu \nu }V_{\mu }V_{\nu }=\left( \rho _{1}+P_{1}\right)
U^{\ast \alpha }U_{\alpha }^{\ast }-\left( P_{1}+P_{2}\right)$,
$\Psi =T^{\mu \nu }\chi _{\mu }\chi _{\nu }=\left( P_{1}+P_{2}\right)
-\left( \rho _{2}+P_{2}\right) W^{\ast \alpha }W_{\alpha }^{\ast }$,
and $\Pi =P_{1}+P_{2}$,
respectively,  the energy-momentum tensor of the two non-interacting perfect fluids
can be written as
\begin{equation}
T^{\mu \nu }=\left( \varepsilon +\Pi \right) V^{\mu }V^{\nu }-\Pi g^{\mu \nu
}+\left( \Psi -\Pi \right) \chi ^{\mu }\chi ^{\nu },  \label{tens}
\end{equation}
where $V^{\mu }V_{\mu }=1=-\chi ^{\mu }\chi _{\mu }$ and $\chi ^{\mu }V_{\mu }=0$ \cite{Le,He}.
The energy-momentum tensor given by Eq.~(\ref{tens}) is the standard form for anisotropic fluids \cite{He}.

The energy density $\varepsilon $ and the radial pressure $\Psi $ are
given by
\begin{widetext}
\begin{equation}
\varepsilon =\frac{1}{2}\left( \rho _{1}+\rho _{2}-P_{1}-P_{2}\right) +\frac{%
1}{2}\sqrt{\left( \rho _{1}+P_{1}+\rho _{2}+P_{2}\right) ^{2}+4\left( \rho
_{1}+P_{1}\right) \left( \rho _{2}+P_{2}\right) \left[ \left( U^{\mu }W_{\mu
}\right) ^{2}-1\right] },  \label{eps}
\end{equation}
\begin{equation}
\Psi =-\frac{1}{2}\left( \rho _{1}+\rho _{2}-P_{1}-P_{2}\right) +\frac{1}{2%
}\sqrt{\left( \rho _{1}+P_{1}-\rho _{2}-P_{2}\right) ^{2}+4\left( \rho
_{1}+P_{1}\right) \left( \rho _{2}+P_{2}\right) \left( U^{\mu }W_{\mu
}\right) ^{2}},  \label{sig}
\end{equation}
\end{widetext}
respectively \cite{Le,He}.

In comoving spherical coordinates $x^{0}=t$, $x^{1}=r$, $%
x^{2}=\vartheta $, and $x^{3}=\phi $ we may choose $V^{1}=V^{2}=V^{3}=0$, $%
V^{0}V_{0}=1$, and $\chi ^{0}=\chi ^{2}=\chi ^{3}=0$, $\chi ^{1}\chi _{1}=-1$%
\cite{Le,He}. Therefore the components of the energy-momentum of two non-interacting
perfect fluids take the form:
$T_{0}^{0}=\varepsilon$, $T_{1}^{1}=-\Psi $, and $T_{2}^{2}=T_{3}^{3}=-\Pi $,
where $\varepsilon $ is the total energy-density of the
mixture of fluids, $\Psi =P_{r}$ is the pressure along the radial
direction, while $\Pi =P_{\perp }$ is the tangential pressure on the $r=$
constant surface.

\section{Anisotropic fluids in spherically symmetric static spacetimes}

In the following we restrict our study of the two-component dark matter to
the static and spherically symmetric case, with the metric  represented as
\begin{equation}\label{metr1}
ds^{2}=e^{\nu (r)}dt^{2}-e^{\lambda (r)}dr^{2}-r^{2}\left( d\vartheta
^{2}+\sin ^{2}\vartheta d\phi ^{2}\right) .
\end{equation}

For the metric given by Eq.~(\ref{metr1}), the Einstein gravitational field
equations, describing the mixture of two  fluids, can be easily integrated to give
\begin{equation}
e^{-\lambda }=1-\frac{2M(r)}{r}, \quad {\rm with} \quad M(r)=4\pi \int \varepsilon r^{2}dr,
\end{equation}
\begin{equation}
\frac{d\nu }{dr}=-2\frac{\Psi ^{\prime }}{\varepsilon +\Psi }+\frac{4}{r}%
\frac{\Pi -\Psi }{\varepsilon +\Psi },  \label{f4}
\end{equation}
\begin{equation}
\frac{dM}{dr}=4\pi \varepsilon r^{2},  \label{eqcont0}
\end{equation}
\begin{equation}
\frac{d\Psi }{dr}=-\frac{\left( \varepsilon +\Psi \right) \left( 4\pi
\Psi r^{3}+M\right) }{r^{2}\left( 1-2M/r\right) }+\frac{2}{r}\left( \Pi
-\Psi \right).  \label{tov0}
\end{equation}
Equation (\ref{f4}) is the consequence of the
conservation of the energy-momentum tensor, $T_{\nu ;\mu }^{\mu }=0$, while Eqs.~(\ref{eqcont0}) and (\ref{tov0}) are the mass continuity, and the hydrostatic equilibrium (TOV) equation.

The galactic rotation curves provide the most direct method of analyzing the
gravitational field inside a spiral galaxy.  The velocity of the cloud, as measured by
an inertial observer far from the source, is given by
\begin{equation}
v^{2}=e^{-\nu }\left[ e^{\lambda }\left( \frac{dr}{dt}\right)
^{2}+r^{2}\left( \frac{d\Omega }{dt}\right) ^{2}\right] .
\end{equation}

For a stable circular orbit $dr/dt=0$, and the tangential velocity of the
test particles can be expressed as
$v_{tg}^{2}=e^{-\nu }r^{2}\left( d\Omega /dt\right) ^{2}$,
where $d\Omega ^2=d\vartheta ^{2}+\sin ^{2}\vartheta d\phi ^{2}$.
In terms of the conserved angular momentum $l=r^{2}\dot{\phi}$ and energy $E=e^{\nu (r)}\dot{t}$,
the tangential velocity is given, for $\theta =\pi /2$, by
$v_{tg}^{2}=e^{\nu }l^2/r^{2}E^2$.
By taking into account the explicit expressions for $l$ and $E$ we obtain
for the tangential velocity of a test particle in a stable circular orbit
the expression $v_{tg}^{2}=r\nu ^{\prime }/2$.

The tangential velocity profile can be obtained as a function of the total
density and of the mass of the dark matter from Eq.~(\ref{f4}) as
\begin{equation}
v_{tg}^{2}(r)=\frac{4\pi \Psi r^{3}+M}{r\left( 1-2M/r\right) }.
\end{equation}

Eq.~(\ref{tov0}), can be written in the following equivalent form
\begin{equation}
\frac{d\Psi }{dr}=-\frac{\left( \varepsilon +\Psi \right) }{r}%
v_{tg}^{2}(r)+\frac{2}{r}\left( \Pi -\Psi \right) .  \label{tovf}
\end{equation}

\section{Dark matter as a mixture of two fluids}

The kinetic energy-momentum tensor $T^{\mu }_{\nu }$ associated to the frozen distribution of dark matter is given by $T^{\mu }_{\nu }=(g/h^3)\int{d^3pf(p)p^{\mu }p_{\nu }/p^0}$, where $f(p)$ is the dark matter particle distribution function, $p^{\mu }$ is the four-momentum,  $\vec{p}$ is the three-momentum, with absolute value $p$,  and $g$ is the number of helicity states, respectively.
The energy density $\epsilon $ of the system is defined as
\begin{equation}
\epsilon =\frac{g}{3h^3}\int{Ef(p)d^3p},
\end{equation}
while the pressure of a system with an isotropic distribution of momenta is given by
\begin{equation}
P=\frac{g}{3h^3}\int{pvf(p)d^3p}=\frac{g}{3h^3}\int{\frac{p^2}{E}f(p)d^3p},
\end{equation}
where the velocity $v$ is related to the momentum by $v=p/E$ \cite{mad}.
In the non-relativistic regime, when $E\approx m$ and $p\approx mv$,   the density $\rho $ of the dark matter is given by $\rho =mn$, where $n$ is the particle number density, while its pressure $P$ can be obtained as \cite{mad}
\begin{equation}\label{pres0}
P=\frac{g}{3h^3}\int{\frac{p^2c^2}{E}f(p)d^3p}\approx 4\pi \frac{g}{3h^3}\int{\frac{p^4}{m}dp},
\end{equation}
giving
\begin{equation}\label{pres1}
P= \sigma ^2\rho ,
\end{equation}
where $\sigma ^2=\langle \vec{v}^{\;2} \rangle /3$, and $\langle \vec{v}^{\;2} \rangle$ is the average squared velocity of the particle. $\sigma $ is the one-dimensional velocity dispersion. In the non--relativistic approximation given by Eqs.~(\ref{pres0}) and (\ref{pres1}), the velocity dispersion $\sigma$  is  a constant only for the
case of the non--degenerate ideal Maxwell--Boltzmann gas, whose Newtonian analogue is
the isothermal sphere. In the following we will consider only the specific case of the ideal gases in the non--relativistic regime, and we will explicitly assume that the dispersion velocity is a constant.

We model the galactic dark matter as a mixture of two fluids, with energy densities $\rho _i, i=1,2$, constant velocity dispersions $\sigma _i,i=1,2$,  pressures $P_i,i=1,2$,  and with four-velocities $U^{\mu }$ and $W^{\mu }$, respectively. Since the two fluids have different four-velocities, we may write $U^{\mu }W_{\mu }=1+b/2$, where generally $b$ is an arbitrary function of the radial coordinate. The functional form of $b$ can be obtained from Eq.~(\ref{alpha}), which gives
\begin{equation}
U^{\mu }W_{\mu }=\frac{1}{2}\tan 2\alpha \left[ \sqrt{\frac{\rho _{1}+p_{1}}{%
\rho _{2}+p_{2}}}-\sqrt{\frac{\rho _{2}+p_{2}}{\rho _{1}+p_{1}}}\right],
\end{equation}
while $b$ can be obtained as
\begin{equation}
b=\tan 2\alpha \left[ \sqrt{\frac{\rho _{1}+p_{1}}{%
\rho _{2}+p_{2}}}-\sqrt{\frac{\rho _{2}+p_{2}}{\rho _{1}+p_{1}}}\right]-2\texttt{}.
\end{equation}

In the following we choose our coordinates such that we are comoving with the fluid with velocity $U^{\mu }$. Thus $U^0=e^{-\nu /2}$ and $U^1=U^2=U^3=0$. We also assume $W^2=W^3=0$. From the conditions $U_{\mu }U^{\mu }=1$ and $W_{\mu }W^{\mu }=1$ we obtain $W^0=(1+b/4)e^{-\nu /2}$ and $W^1=\sqrt{(b/2)(1+b/8)}e^{-\lambda /2}$.
By assuming that the physical parameters of the two fluids satisfy the
condition
\begin{equation}
2b\left( 1+\frac{b}{8}\right) \frac{\left( \rho _{1}+P_{1}\right) \left(
\rho _{2}+P_{2}\right) }{\left( \rho _{1}+\rho _{2}+P_{1}+P_{2}\right) ^{2}}%
<<1,
\end{equation}
the energy density, and the radial and tangential pressures can be obtained
as
\begin{equation}
\varepsilon =\rho _{1}+\rho _{2}+\frac{b}{2}\left( 1+\frac{b}{8}\right)
\frac{\left( \rho _{1}+P_{1}\right) \left( \rho _{2}+P_{2}\right) }{\left(
\rho _{1}+\rho _{2}+p_{1}+p_{2}\right) },
\end{equation}
\begin{equation}
\Psi =P_{1}+P_{2}+\frac{b}{2}\left( 1+\frac{b}{8}\right) \frac{\left( \rho
_{1}+P_{1}\right) \left( \rho _{2}+P_{2}\right) }{\left( \rho _{1}+\rho
_{2}+P_{1}+P_{2}\right) },
\end{equation}
\begin{equation}
\Pi =P_{1}+P_{2}.
\end{equation}
In the following we assume that the masses $m_1$ and $m_2$ and the number densities $n_1$ and $n_2$ of the dark matter particles satisfy the scaling relation $m_1/m_2=n_2/n_1$, giving $\rho _{1}=\rho _{2}=\rho $ for the densities of the two components. Taking into account that  $P_{1}=\sigma _{1}^{2}\rho _{1}$, $P_{2}=\sigma
_{2}^{2}\rho _{2}$, the expression of $b$ can be written as
\begin{equation}
b=\tan 2\alpha \left( \frac{\sigma _{1}^{2}-\sigma _{2}^{2}}{%
\sqrt{1+\sigma _{1}^{2}}\sqrt{1+\sigma _{2}^{2}}}\right) -2.
\end{equation}
Hence for this particular case it follows that $b={\rm constant}$. Moreover,
since the rotation angle is arbitrary, for $\sigma _1>\sigma _2$, we can always find an angle $\alpha $
so that the condition $b>0$ is also satisfied. In the following we will restrict our analysis to the range of positive values of $b$.

\begin{figure}[!ht]
\includegraphics[width=0.98\linewidth]{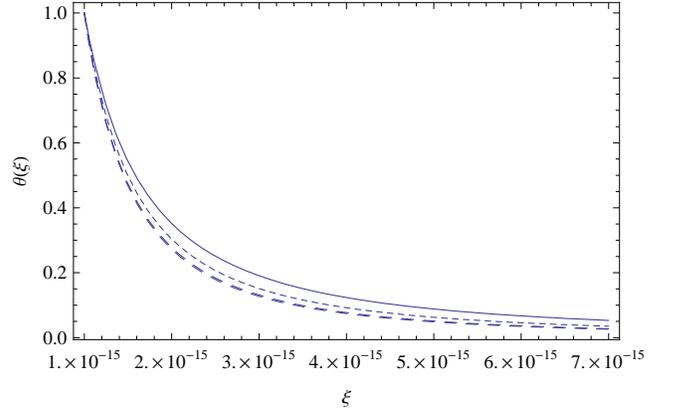}
\caption{Dimensionless density profile of a two-component dark matter fluid, with $\sigma _1=950$ km/s and $\sigma _2=750$ km/s, for different values of $b$:
$b=0.0001$ (solid curve), $b=0.0002$ (dotted curve), $b=0.0004$ (dashed curve), and $b=0.0005$ (long-dashed curve), respectively.}
\label{fig1}
\end{figure}

\begin{figure}[!ht]
\includegraphics[width=0.98\linewidth]{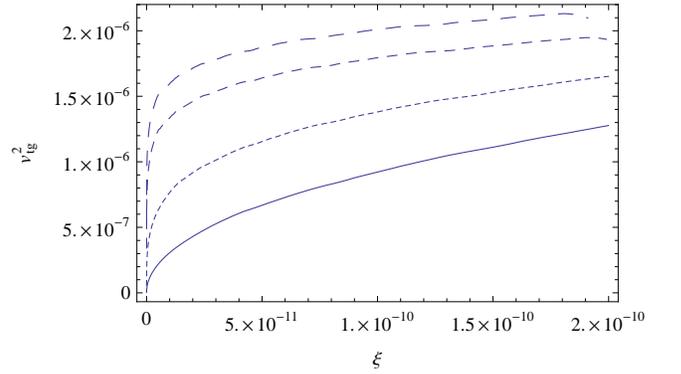}
\caption{Tangential velocities of test bodies in a two-component dark matter fluid, with $\sigma _1=950$ km/s and $\sigma _2=750$ km/s, for different values of $b$:
$b=0.0001$ (solid curve), $b=0.0002$ (dotted curve), $b=0.0004$ (dashed curve), and $b=0.0005$ (long-dashed curve), respectively.}
\label{fig2}
\end{figure}

Therefore the energy
density and the pressures of the two fluid dark matter mixture can be
written as $\varepsilon =\varepsilon _{0}\rho $, $ \Psi =\Psi _{0}\rho $,
$\Pi =\Pi _{0}\rho $, where
$\varepsilon _{0}=\left[ 2+\left( b/2\right)
\left( 1+b/8\right) \right]$,
$\Psi _{0}=\left[ \sigma _{1}^{2}+\sigma
_{2}^{2}+\left( b/2\right) \left( 1+b/8\right) \right]$,
and $\Pi
_{0}=\sigma _{1}^{2}+\sigma _{2}^{2}$,
respectively.  By introducing a set of dimensionless
variables $\left(\theta , \xi, m\right)$, defined as
$\varepsilon =\varepsilon _{c}\theta $, $r=\xi /\sqrt{4\pi
\varepsilon _{c}}$, $M=m/\sqrt{4\pi\varepsilon _{c}}$,
where $\varepsilon _{c}$ is the central density of the dark matter halo, the
continuity equation, the TOV equation and the tangential velocity can be
written in a dimensionless form as
\begin{equation}
\frac{dm}{d\xi }=\xi ^{2}\theta ,  \label{dim1}
\end{equation}
\begin{equation}
\frac{d\theta }{d\xi }=-\frac{\left( 1+\gamma _{\varepsilon }\right) \theta
\left( \xi ^{3}\theta /\gamma _{\varepsilon } +m\right) }{\xi ^{2}\left( 1-2m/\xi \right) }%
+2\left( \gamma _{\Pi }-1\right) \frac{\theta }{\xi },  \label{dim2}
\end{equation}
\begin{equation}
v_{tg}^{2}=\frac{\xi ^3\theta /\gamma _{\varepsilon } +m}{\xi \left( 1-2m/\xi \right) },
\end{equation}
where $\gamma _{\varepsilon }=\varepsilon _{0}/\Psi _{0}$ and $\gamma _{\Pi
}=\Pi _{0}/\Psi _{0}$, respectively. Equations (\ref{dim1}) and (\ref{dim2}) must be integrated with the boundary conditions $\theta \left( 0\right) =1$ and $m(0)=1$. The variation of the two-fluid dark matter density profile and of the tangential velocity of test bodies is represented, for fixed values of $\sigma _1$ and $\sigma _2$, in Figs.~\ref{fig1} and \ref{fig2}.

\section{Cosmological implications}

Since the two-fluid dark matter mixture represents an anisotropic
fluid, on a cosmological scale the corresponding geometry is also
anisotropic. In a homogenous universe a two-fluid mixture can be described
by a flat Bianchi type I geometry, with the line element given by
\begin{equation}
ds^{2}=dt^{2}-a_{1}^{2}(t)dx^{2}-a_{2}^{2}(t)dy^{2}-a_{3}^{2}(t)dz^{2},
\label{7}
\end{equation}%
where $a_i(t),i=1,2,3$ are the directional scale factors. In this geometry the gravitational field equations take the form
\begin{equation}
3\dot{H}+H_{1}^{2}+H_{2}^{2}+H_{3}^{2}=-\frac{1}{2}\left( \varepsilon +\Psi
+2\Pi \right) ,  \label{8}
\end{equation}%
\begin{eqnarray}
\frac{1}{V}\frac{d}{dt}\left( VH_{1}\right)  &=&\frac{1}{2}\left(
\varepsilon -\Psi \right) ,  \label{9} \\
\frac{1}{V}\frac{d}{dt}\left( VH_{2}\right)  &=&\frac{1}{V}\frac{d}{dt}%
\left( VH_{3}\right) =\frac{1}{2}\left( \varepsilon -\Pi \right) ,
\end{eqnarray}%
where we have denoted $V=a_{1}a_{2}a_{3}$ , $H_{i}=\dot{a}_{i}/a_{i},i=1,2,3$ and
$H=\frac{1}{3}\sum_{i=1}^{3}H_{i}=\dot{V}/3V$, respectively. On a cosmological
scale the pressures of the two-fluid dark matter obey the conditions $\Psi
,\Pi \ll\varepsilon $, and hence in the gravitational field equations we can neglect $\Psi $ and $\Pi $ with respect to $\varepsilon $.  By adding Eqs.~(\ref{9}) we obtain
\begin{equation}
\dot{H}+3H^{2}=\frac{1}{2}\varepsilon ,  \label{10}
\end{equation}%
leading to $H_{i}=H+K_{i}/V,i=1,2,3,$ where $K_{i},i=1,2,3$ are constants of
integration satisfying the consistency condition $\sum_{i=1}^{3}K_{i}=0$. We
also denote $K^{2}=\sum_{i=1}^{3}K_{i}^{2}.$ From the conservation of the
energy density of the dark matter, $\dot{\varepsilon}+3H\varepsilon =0$, we
obtain $\varepsilon =\varepsilon _{C}/V$, where $\varepsilon _{C}$ is an
arbitrary integration constant. Eq.~(\ref{10}) takes the form $\ddot{V}%
=(3/2)\varepsilon _{C}$, with the general solution given by $%
V(t)=(3/4)\varepsilon _{C}t^{2}+C_{1}t+C_{2}$, with $C_{i},i=1,2$ arbitrary
integration constant. The large-time behavior of an anisotropic
cosmological model can be obtained from the study of the  anisotropy
parameter $A$, defined by
\begin{equation}
A=\frac{1}{3}\sum_{i=1}^{3}\left( \frac{H_{i}-H}{H}\right) ^{2}=\frac{3K^{2}%
}{\dot{V}^{2}}.
\end{equation}

If $A=0$ the cosmological model is isotropic. For the two-fluid dark matter
model  $A=3K^{2}/\left[ (3/2)\varepsilon _{C}t+C_{1}\right] ^{2}$, and
therefore in the large-time limit $A\rightarrow 0$. Thus, on a cosmological
scale, and in the large time limit,  the initially anisotropic dark matter fluid mixture behaves like an
isotropic fluid.

\section{Discussions and final remarks}

In the present paper we have considered the theoretical possibility that dark matter may be modeled as a mixture of two non-interacting perfect fluids, with different four-velocities. This model is formally equivalent to a single anisotropic fluid. By considering a purely kinetic description of dark matter the basic equations describing the dark matter mixture have been obtained, and integrated numerically. In the particular case in which $\Psi, \Pi \ll\varepsilon$,  we can approximate $\Psi =\Pi \approx 0$, corresponding to two non-interacting pressureless fluids. For this case the theoretical model reduces to a single anisotropic fluid with vanishing tangential pressure. In the general case of arbitrary fluids there is a general $r$-dependent functional relationship between the energy density and the radial and tangential pressures, which differs radically from the simple barotropic equation of state. On the other hand, in was shown in \cite{Sax} that half of the 14 galaxies considered in the study are well fit by the polytropic halo model, despite its serious physical simplifications. The necessity of considering  non-standard dark matter models with pressure is justified by our uncertainties in the knowledge about the nature of the dark matter particle, as well as by the fact that this model gives a much better description of the observational results, as compared to the pressureless case. As a consistency check of our model we have considered the cosmological behavior of a mixture of two fluids, and we have showed that such a mixtures always isotropizes in the large time limit.

The approach used in the present paper is different from the method used in \cite{Haag1} for the study of dust shells. First of all, the two fluids are described as a single effective anisotropic fluid.  While in \cite{Haag1} a non-comoving frame of reference is adopted, due to our transformation of the energy-momentum tensor, we can write down and study the field equations in a comoving frame. Moreover, the inclusion of the pressure terms is straightforward. We have also to point out the differences in geometry: while in \cite{Haag1} the time-dependent Lemaitre-Tolman model is used, and additional symmetries are imposed in order to reduce the number of
variables, in the present paper we are considering the properties of dark matter in a spherically symmetric static geometry. The present model can be easily generalized for other equations of state of dark matter, or for cases involving the presence of condensates or scalar fields.
These possibilities, and their impact on dark matter distribution, as well as some cosmological implications, will be considered in some future studies.

\section*{Acknowledgments}

The work of TH was supported by a GRF grant of the government of the Hong Kong SAR. FSNL acknowledges financial support of the Funda\c{c}\~{a}o para a Ci\^{e}ncia e Tecnologia through the grants PTDC/FIS/102742/2008 and CERN/FP/116398/2010.

\end{document}